\documentclass[twocolumn,prl,showpacs]{revtex4}
\usepackage{graphicx}
\begin{document}
\topmargin 0 cm

\title{ 
{\boldmath Dalitz Plot Analysis of the Decay $D^+\to K^-\pi^+\pi^+$ and Indication
 of \\ a Low-Mass Scalar $K\pi$ Resonance}
}

\author{
    E.~M.~Aitala,$^{9}$
       S.~Amato,$^1$
    J.~C.~Anjos,$^1$
    J.~A.~Appel,$^{5}$
       D.~Ashery,$^{16}$
       S.~Banerjee,$^5$
       I.~Bediaga,$^1$
       G.~Blaylock,$^8$
    S.~B.~Bracker,$^{17}$
    P.~R.~Burchat,$^{15}$
    R.~A.~Burnstein,$^6$
       T.~Carter,$^5$
    H.~S.~Carvalho,$^{1}$
    N.~K.~Copty,$^{14}$
    L.~M.~Cremaldi,$^{9}$
       C.~Darling,$^{20}$
       K.~Denisenko,$^5$
       S.~Devmal,$^3$
       A.~Fernandez,$^{11}$
    G.~F.~Fox,$^{14}$
       P.~Gagnon,$^2$
       C.~G\"obel,$^{1,12}$
       K.~Gounder,$^{9}$
    A.~M.~Halling,$^5$
       G.~Herrera,$^4$
       G.~Hurvits,$^{16}$
       C.~James,$^5$
    P.~A.~Kasper,$^6$
       S.~Kwan,$^5$
    D.~C.~Langs,$^{14}$
       J.~Leslie,$^2$
       B.~Lundberg,$^5$
       J.~Magnin,$^1$       
       A.~Massafferri,$^1$
       S.~MayTal-Beck,$^{16}$
       B.~Meadows,$^3$
 J.~R.~T.~de~Mello~Neto,$^1$
       D.~Mihalcea,$^7$
    R.~H.~Milburn,$^{18}$
    J.~M.~de~Miranda,$^1$
       A.~Napier,$^{18}$
       A.~Nguyen,$^7$
    A.~B.~d'Oliveira,$^{3,11}$
       K.~O'Shaughnessy,$^2$
    K.~C.~Peng,$^6$
    L.~P.~Perera,$^3$
    M.~V.~Purohit,$^{14}$
       B.~Quinn,$^{9}$
       S.~Radeztsky,$^{19}$
       A.~Rafatian,$^{9}$
    N.~W.~Reay,$^7$
    J.~J.~Reidy,$^{9}$
    A.~C.~dos Reis,$^1$
    H.~A.~Rubin,$^6$
    D.~A.~Sanders,$^{9}$
 A.~K.~S.~Santha,$^3$
 A.~F.~S.~Santoro,$^1$
       A.~J.~Schwartz,$^{3}$
       M.~Sheaff,$^{19}$
    R.~A.~Sidwell,$^6$
    A.~J.~Slaughter,$^{20}$
    M.~D.~Sokoloff,$^3$
    C.~J.~Solano~Salinas,$^{1,13}$
    N.~R.~Stanton,$^7$
    R.~J.~Stefanski,$^5$  
       K.~Stenson,$^{19}$ 
    D.~J.~Summers,$^{9}$
       S.~Takach,$^{20}$
       K.~Thorne,$^5$
    A.~K.~Tripathi,$^{7}$
       S.~Watanabe,$^{19}$
 R.~Weiss-Babai,$^{16}$
       J.~Wiener,$^{10}$
       N.~Witchey,$^7$
       E.~Wolin,$^{20}$
    S.~M.~Yang,$^7$
       D.~Yi,$^{9}$
       S.~Yoshida,$^7$
       R.~Zaliznyak,$^{15}$ and
       C.~Zhang$^7$ \\  
    ~~   \\  
       (Fermilab E791 Collaboration) \\
  ~~\\
}

\affiliation{
$^1$Centro Brasileiro de Pesquisas F{\'\i}sicas, Rio de Janeiro, Brazil \\
$^2$University of California, Santa Cruz, California 95064 \\
$^3$University of Cincinnati, Cincinnati, Ohio 45221\\
$^4$CINVESTAV, Mexico City, Mexico\\
$^5$Fermilab, Batavia, Illinois 60510\\
$^6$Illinois Institute of Technology, Chicago, Illinois 60616\\
$^7$Kansas State University, Manhattan, Kansas 66506\\
$^8$University of Massachusetts, Amherst, Massachusetts 01003\\
$^9$University of Mississippi-Oxford, University, Mississippi 38677\\
$^{10}$Princeton University, Princeton, New Jersey 08544\\
$^{11}$Universidad Autonoma de Puebla, Puebla, Mexico\\
$^{12}$Universidad de la Rep\'ublica, Montevideo, Uruguay\\
$^{13}$Universidade Federal de Itajub\'a, Itajub\'a, Brazil \\
$^{14}$University of South Carolina, Columbia, South Carolina 29208\\
$^{15}$Stanford University, Stanford, California 94305\\
$^{16}$Tel Aviv University, Tel Aviv, Israel\\
$^{17}$Box 1290, Enderby, British Columbia, V0E 1V0, Canada\\
$^{18}$Tufts University, Medford, Massachusetts 02155\\
$^{19}$University of Wisconsin, Madison, Wisconsin 53706\\
$^{20}$Yale University, New Haven, Connecticut 06511
}


\begin{abstract}
We study the Dalitz plot of the decay $D^+\to K^-\pi^+\pi^+$  with a sample of 15090 events from
Fermilab experiment E791. Modeling the decay amplitude as the coherent sum of known $ K \pi $ 
resonances and a uniform nonresonant term, we do not obtain an acceptable fit. 
If we allow the mass and width of the $K^*_0(1430)$ to float, we obtain values
consistent with those from PDG but the $\chi^2$ per degree of freedom of the fit is 
still unsatisfactory. A good fit is found when we allow for the 
presence of an additional scalar resonance, with mass $797\pm 19\pm 43$~MeV/c$^2$ and 
width $410\pm 43\pm 87$~MeV/c$^2$. The mass and width of the $K^*_0(1430)$ become 
$1459\pm 7\pm 5$~MeV/c$^2$ and $175\pm 12\pm 12$~MeV/c$^2$, respectively. 
Our results provide new information on the scalar sector in hadron spectroscopy. 
\end{abstract}
\maketitle

In this paper we present a Dalitz plot analysis of the Cabibbo-favored 
decay $D^+\to K^-\pi^+\pi^+$ using data from Fermilab experiment E791. 
Previous analyses of this decay \cite{e691-kpipi,e687-kpipi}
modeled the amplitude as the coherent sum of known $ K \pi $ 
resonances and a uniform non-resonant (NR) term. They observed that 
the NR term is strongly dominant, unlike other $ D $ 
decays, and that the sum of the decay fractions substantially 
exceeds unity, indicating large interference. 
Moreover, the fits did not describe the Dalitz plot distributions 
well. In our analysis, we obtain similar results but
with higher statistics. We study variations in the underlying model, 
including changes in form factors, tuning of resonance parameters, and the addition
of known and new resonance structures. 
In particular, we investigate the scalar sector for which
there has been much uncertainty for many years.

This study is based on the Fermilab E791 sample of $ 2 \times 10^{10} $ events 
produced from interactions of a 500 GeV/c $\pi^-$ beam with five thin target foils 
(one platinum, four diamond). Descriptions of the detector, data set, reconstruction, 
and vertex resolutions can be found in Ref.~\cite{ref791}.
A clean sample of $K^-\pi^+\pi^+$ decays (charge-conjugate modes are implicit 
throughout this paper) was selected by requiring that the 3-prong decay (secondary) 
vertex be well-separated from the production (primary) vertex and located 
outside any solid material. 
The sum of the momentum vectors of the three tracks from the 
secondary vertex was required to point to the primary vertex, 
and each of the three tracks was required to pass closer to 
the secondary vertex than to the primary. 
We restricted the $p_T^2$ and $x_F$ ranges of the $D^+$ candidates to ensure 
an accurate model of our experiment in the Monte Carlo (MC) simulation. 
Finally, we required that the odd-charge track (track with charge 
opposite that of the $D^\pm$ candidate) from the secondary vertex be 
consistent with kaon identification in the \v{C}erenkov counters~\cite{cerenkov}.

\begin{figure}
\includegraphics[width=9cm]{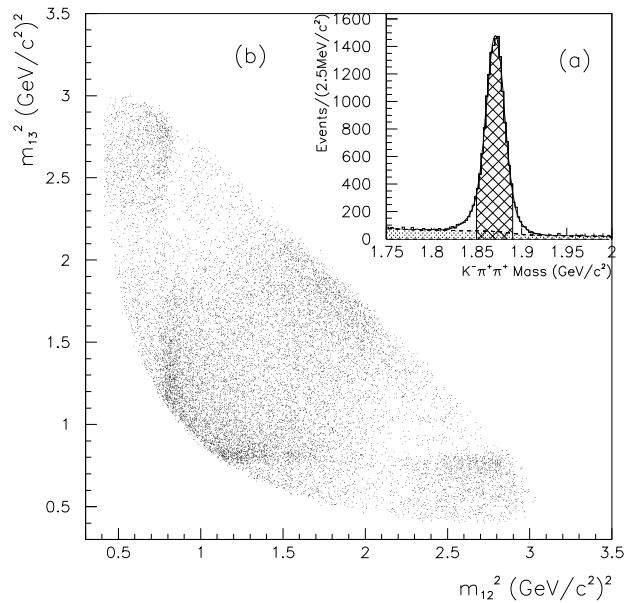}
\caption {\label{fig1} (a) The $K^-\pi^+\pi^+$ invariant mass spectrum. The shaded area 
corresponds to the background level. The crosshatched region is the sample 
used for the Dalitz plot analysis. (b) The $D^+\to K^-\pi^+\pi^+$ Dalitz plot, 
symmetrized for the indistinguishable pions. }
\end{figure}

We fit the $K^-\pi^+\pi^+ $ invariant mass distribution shown in 
Fig.~\ref{fig1}(a) by the sum of $D^+$ signal and background terms. 
The signal was represented by the sum of two Gaussians, with parameters 
determined by the fit. We used MC simulations and data to determine 
both the shape and the size of charm backgrounds. The significant 
sources are reflections from $D_s^+\to K^-K^+\pi^+$ (via $\bar K^*K^+$ 
and $\phi\pi^+$), in which one kaon is misidentified 
as a pion. Other sources of charm background are either 
negligible or broadly distributed and thus safely included when we 
estimate combinatorial background, which 
was represented by an exponential function.  
The number of $D^+$ 
candidates obtained from the fit is $16190\pm 139$.

For the Dalitz plot analysis, we selected candidates in the 
$K\pi\pi$ mass range 1.85--1.89 GeV/c$^2$ (crosshatched 
region in Fig.~\ref{fig1}(a)). This results in 15090 events, with about 
6\% due to background. Fig.~\ref{fig1}(b) shows the corresponding Dalitz 
plot, $m^2_{12}$ vs. $m^2_{13}$, in which the kaon candidate is labeled 
particle 1, and the plot is symmetrized with respect to the two pions 
(particles 2, 3).

To study the resonant structure in Fig.~\ref{fig1}(b), an unbinned maximum 
likelihood fit is used. 
The likelihood ${\cal L}$ is computed as ${\cal L} = \prod_{\rm events} 
\left[\sum_{i=1}^3n_{B_i}{\cal P}_{B_i}+ n_S{\cal P}_S\right]$,
where ${\cal P}_{B_i}$ and ${\cal P}_S$ are the normalized probability 
density functions (PDF's) for background and signal, respectively, 
and $n_{B_i}$ and $n_S$ are their fractional contributions.
Each background PDF is written as ${\cal P}_{B_i} = {1\over N_{B_i}} b_i(M) 
{\cal F}_{B_i}(m^2_{12},m^2_{13})$, where $N_{B_i}$ is the normalization, 
$b_i(M)$ is the distribution in the $K\pi\pi$ mass spectrum, and 
${\cal F}_{B_i}$ is the shape in the Dalitz plot. The shape of the 
combinatorial background is obtained from a fit to events above the 
signal peak in the $K\pi\pi$ mass range 1.92--1.96 GeV/c$^2$. The 
shapes of the $D_s^+\to\bar K^*K^+$ and $D_s^+\to\phi\pi^+$ backgrounds 
are from MC simulations. 

The signal PDF is ${\cal P}_S = {1\over N_S} g(M) \varepsilon(m^2_{12},m^2_{13}) 
|{\cal A}|^2$, where $N_S$ is the normalization, $g(M)$ describes the signal 
shape in the $K\pi\pi$ mass spectrum, and $\varepsilon(m^2_{12},m^2_{13})$ 
is the acceptance across the Dalitz plot, 
including smearing. The signal amplitude $\cal A$ is a 
coherent sum of a uniform NR amplitude and resonant $K \pi $ amplitudes,
\begin{equation}
{\cal A} = a_0e^{i\delta_0}{\cal A}_0 + \sum_{n=1}^N a_ne^{i\delta_n}
{\cal A}_n(m^2_{12},m^2_{13})\ ,
\label{ampl}
\end{equation} 
\noindent
where each term is Bose symmetrized for the pions: 
${\cal A}_n = {\cal A}_n[({\bf 12}){\bf3}] + {\cal A}_n[({\bf 13}){\bf 2}]$.
The coefficients $a_n$ are magnitudes and the $\delta_n$ are 
relative phases.

Our first fit, referred to as Model A, includes only well-established 
resonances and fixes their masses and widths to PDG \cite{pdg} values. 
This approach has been used in previous Dalitz-plot analyses 
({\it e.g.}, Refs.~\cite{e687-kpipi,argus}). The NR amplitude ${\cal A}_0$ is 
represented by a constant; {\it i.e.}, it has no magnitude or phase 
variation across the Dalitz plot.
Each resonant amplitude ${\cal A}_n$ ($n >0$) is written as
${\cal A}_n = \ BW_n~F_D^{(J)}~ F_n^{(J)}~ {\cal M}_n^{(J)}$.
The $BW_n$ factor is the relativistic Breit-Wigner propagator, 
$BW_n=\{m^2_n - m^2 -im_n\Gamma_n(m)\}^{-1}$,
where $m$ is the invariant mass of the $K\pi$ pair forming a resonance 
(either $m_{12}$ or $m_{13}$), $m_n$ is the resonance mass, and $\Gamma_n(m)$ 
is the mass-dependent width. 
The factors $F_D^{(J)}$ and $F_n^{(J)}$ are Blatt-Weisskopf penetration factors \cite{blatt}, 
which depend on the spin $J$ and the radii of the relevant mesons. 
In Model A, the radii are fixed as $r_D=5$~GeV$^{-1}$ for the $D$ meson 
and $r_R=1.5$~GeV$^{-1}$ for all $K\pi$ resonances~\cite{argus}. 
No form factors $F$ are used for scalar resonances. The term
${\cal M}_n^{(J)}$ accounts for the decay angular distribution. 
Ref.~\cite{d3pi} gives detailed expressions for all these functions; note that  
we use the opposite sign for the $BW_n$ term, for easier 
comparison of our results to those of Ref.~\cite{e687-kpipi}.

For Model A, we fix the NR parameters to be $a_0=1$ and $\delta_0=0$, 
and include all well-established
$K\pi$ resonances; the only free parameters of the fit are the magnitudes 
$a_n$ and phases $\delta_n$ of the resonances.
The so-called decay fraction for each mode is obtained by integrating 
its intensity (squared amplitude) over the Dalitz plot and dividing by 
the integrated intensity with all modes present. The fit results 
are listed in Table~\ref{tab1}. We observe 
contributions from the same channels reported previously \cite{e691-kpipi,e687-kpipi}; 
{\it i.e.}, a high NR decay fraction (over 90\%), followed by 
$\bar K^*_0(1430)\pi^+$, $\bar K^*(892)\pi^+$, and $\bar K^*(1680)\pi^+$. 
We also measure a small but statistically significant contribution from 
$\bar K^*_2(1430)\pi^+$. 
No other resonances considered are found to contribute.
The sum of the decay fractions is $\sim 140\%$, indicating 
a high level of interference.

\begin{figure}[t]
\includegraphics[width=9cm]{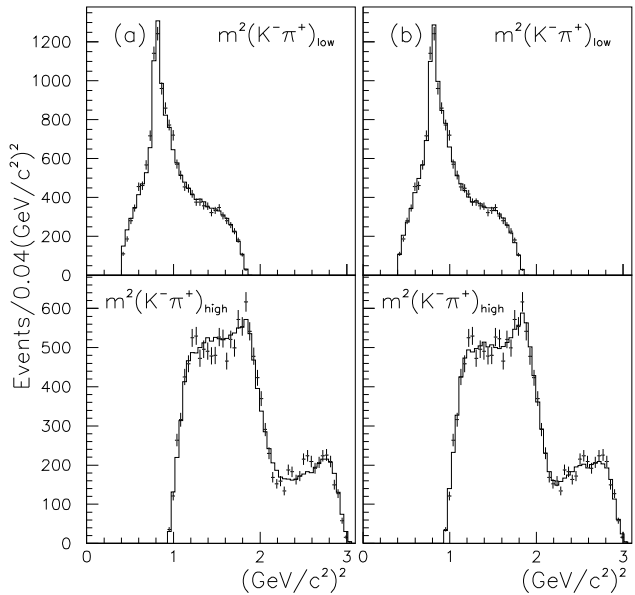}
\caption {\label{fig2} $m^2(K\pi)_{\rm low}$ and $m^2(K\pi)_{\rm high}$ projections for data
(error bars) and fast MC (solid line): Models A (a) and C (b).}
\end{figure}

To assess the quality of the fit, we developed a fast-MC algorithm 
which produces binned Dalitz plot densities according to signal and 
background PDF's, including detector efficiency and resolution. 
A $\chi^2$ is calculated from the difference between the binned 
Dalitz-plot-density distribution for data and that for fast-MC events 
generated using the parameters obtained from the fit of Model~A\@. The 
$ \chi^2 $ summed over all bins is 167 for 63 degrees of freedom~($ \nu $).
The largest contributions to this $\chi^2$ come from bins at low $K\pi$ mass.
In Fig.~\ref{fig2}(a) we show the mass-squared projections; the top (bottom) 
plot shows the lower (higher) mass combination. The points represent data and 
the solid line represents fast-MC simulation of Model A\@. 
The main discrepancies 
occur below 0.6~(GeV/c$^2$)$^2$ and around 2.5~(GeV/c$^2$)$^2$. 
These discrepancies, and the large value of $ \chi^2 / \nu $,
motivated us to study alternative ways to model the decay amplitude.

\begin{table}
\protect\caption{
Results of the Dalitz plot fits. Models A and B are without 
$\kappa$; Model C is with $\kappa$. For each mode the first row lists the
decay fraction in percent, the second row lists the magnitude of the amplitude
($a^{}_n$), and the third row lists the relative phase ($\delta^{}_n$).
The first error listed is statistical, and the second error (when listed) is systematic.
}
\centering
 \begin{tabular}{c c c c}\hline \hline 
 Mode                 &{\bf Model A}  & {\bf Model B} &  {\bf Model C}   \\
\hline 
NR            & $90.9\pm 2.6$ & $89.5\pm 16.1$ & $13.0\pm 5.8\pm 4.4$   \\ 
              & $1.0$ (fixed) & $2.72\pm 0.55$ &$1.03\pm 0.30\pm 0.16$ \\
              & $0^\circ$(fixed)  & $(-49\pm 3)^\circ$ & $(-11\pm 14\pm 8)^\circ$ \\
\hline
                     & --    & -- & $47.8\pm 12.1\pm 5.3$ \\
$\kappa\pi^+$        & --    & -- & $1.97 \pm 0.35\pm 0.11$ \\
		     & --    & -- & $(187\pm 8\pm 18)^\circ$ \\
\hline		     
                     & $13.8\pm 0.5$  & $12.1\pm 3.3$ & $12.3\pm 1.0\pm 0.9$ \\
$\bar K^*(892)\pi^+$ & $0.39\pm 0.01$ & 1.0 (fixed) & $1.0$ (fixed) \\
                     & $(54\pm 2)^\circ$ & $0^\circ$ (fixed) &  $0^\circ$ (fixed) \\
\hline
                        & $30.6\pm 1.6$ & $28.7\pm 10.2$ & $12.5\pm 1.4\pm 0.5$ \\
$\bar K^*_0(1430)\pi^+$ & $0.58\pm 0.01$ & $1.54\pm 0.75$ & $1.01 \pm 0.10\pm 0.08$ \\
                        & $(54\pm 2)^\circ$ & $(6\pm 12)^\circ$ & $(48\pm 7\pm 10)^\circ$ \\
\hline
                        & $0.4\pm 0.1$ & $0.5\pm 0.3$ & $0.5\pm 0.1\pm 0.2$\\
$\bar K^*_2(1430)\pi^+$ & $0.07\pm 0.01$ &  $0.21\pm 0.18$ & $0.20 \pm 0.05\pm 0.04$ \\
                        & $(33\pm 8)^\circ$ & $(-3\pm 26)^\circ$ & $(-54\pm 8\pm 7)^\circ$ \\
\hline
                        & $3.2\pm 0.3$ & $3.7\pm 1.9$ & $2.5\pm 0.7\pm 0.3$ \\
$\bar K^*(1680)\pi^+$   & \,$0.19\pm 0.01$ ~& ~~$0.56\pm 0.48$ ~~& ~$0.45 \pm 0.16\pm 0.02$\, \\
                        & $(66\pm 3)^\circ$ & $(36\pm 25)^\circ$ & $(28\pm 13\pm 15)^\circ$ \\
\hline

$\chi^2/\nu$            & 167/63 & 126/63 & 46/63 \\ \hline \hline			
\end{tabular}
\label{tab1}
\end{table}

For our second fit, Model B, we allow the mass and
width of the scalar $ K^*_0(1430) $ resonance to float. 
In addition, we include form factors to account for the finite size of the decaying 
mesons in this scalar transition
\cite{torn,cleotau}.
The amplitude is written as 
$F^{(0)}_DF^{(0)}_nBW_n$, in which the form factors are Gaussian:
$F^{(0)}=\exp(-p^*{}^2 / (2k_0^2))$. The factor $p^*$ is the 
momentum of the decay products, $k_0=\sqrt{6}/r$, and $r$ is the 
decaying meson radius. These radii ($r_D$ and $r_R$ introduced above)
become additional free parameters in the fit.  The results of this fit are 
listed in the middle column of Table~\ref{tab1}. The decay fractions 
obtained are very similar to those found for Model~A, but the 
$\chi^2/\nu$ is improved, dropping from 167/63 to 126/63. The mass 
and width of the $K^*_0(1430)$ obtained by the fit
are $1416\pm 27$~MeV/c$^2$ and $250\pm 21$~MeV/c$^2$ 
respectively, which are close to the PDG values of $1412\pm 6$~MeV/c$^2$ and 
$294\pm 23$~MeV/c$^2$ \cite{pdg}. The meson radii obtained are 
$r_D = 0.8\pm 1.0$~GeV$^{-1}$ and $r_R=1.8\pm~3.4$~GeV$^{-1}$.

Since Model B still does not give a satisfactory fit, we allow 
for an additional scalar amplitude (Model~C). 
For this extra amplitude, we use Gaussian form 
factors similar to those used for the $K^*_0(1430)$ \cite{thresh}. 
The decay fractions and relative phases 
obtained by the fit  are listed 
in the right-most column of Table~\ref{tab1}. In the table we denote 
the additional scalar resonance as ``$\kappa$''. 
In fact, discussions of the existence of such a resonance are found
in the literature \cite{kapparef1,nokapparef}.
The fit results are 
very different from those obtained for Models A and B, 
the NR decay fraction drops from 90\% to $(13\pm 6)$\%; the $\kappa\pi^+$ 
channel is dominant with a decay fraction of $(48\pm 12)$; and
the sum of all fractions is $\sim 90\%$, with smaller interference 
effects. 
The $\chi^2/\nu$  decreases to 46/63, substantially lower than those for Models A and B\@.
For the $K_0^*(1430)$ resonance, we measure $m_{K_0^*(1430)}=1459\pm 7\pm 
12$~MeV/c$^2$ and $\Gamma_{K_0^*(1430)}=175\pm 12\pm 12$~MeV/c$^2$. These 
values are significantly higher and narrower, respectively, than those 
given by the PDG~\cite{pdg} which are taken from LASS~\cite{lass}. See also~\cite{dunwoodie}.
The mass and width of the additional resonance ($\kappa$) are
$797\pm 19\pm 43$ MeV/c$^2$ and $410\pm 43\pm 87$ MeV/c$^2$, respectively.
The meson radii obtained in Model~C are 
$r_D = 5.0\pm 0.5$~GeV$^{-1}$ and $r_R = 1.6\pm 1.3$~GeV$^{-1}$.
The $K\pi$ mass-squared projections are shown in Fig.~\ref{fig2}(b).

To better understand our results for Model~C, we perform the following 
test. For Models~B and~C we use the fast-MC to generate an ensemble of 
1000 ``experiments,'' with each experiment having a sample size 
Poisson-distributed around our observed sample size. For each 
experiment we calculate
$ \Delta w_{B,C} \equiv -2 ( {\rm ln}\, {\cal L} _B - {\rm ln} \, {\cal L} _{C} )$, 
where $ {\cal L} _{B}$ and ${\cal L} _{C}$ are the likelihood functions 
evaluated with parameters from Models~B and~C, respectively. 
For the ensemble 
generated without $\kappa\pi^+$, $ \langle \Delta w_{B,C}\rangle = -123 $; 
{\it i.e.}, Model B has greater likelihood. For the ensemble generated {\it with\/} 
$\kappa\pi^+$,  $ \langle \Delta w_{B,C} \rangle = 143 $; {\it i.e.}, Model C has 
greater likelihood. In both cases the rms
of the distributions is about~23. For the data, $ \Delta w_{B,C} = 123 $. 
This value is similar to that obtained for fast-MC events generated 
according to Model~C, and it is very different from that of events generated 
according to Model~B\@.

We investigate the stability of our results and estimate systematic 
errors from the following studies. We divide the total sample into disjoint subsamples
according to $D$ charge, bins of $p_T^2$, $x_F$, and $K\pi\pi$ invariant mass, and
repeat the analysis. 
We perform fit variations by changing fixed parameters of the fit:
background parameterizations, and the mass and width of the $K^*(1680)$.
We also repeat the analysis for samples selected with tighter and
looser event selection criteria. These studies lead to the 
systematic errors quoted in the text and Table \ref{tab1}.
The mass obtained for the $\kappa$, and the mass and width obtained 
for the $K^*_0(1430)$, are found to vary relatively little; {\it e.g.}, $m^{}_\kappa$ 
varies in the range 770--860~MeV/$c^2$. 
The width of the $\kappa$, and the $\kappa\pi^+$ and NR decay fractions,
are found to vary much more: $\Gamma^{}_\kappa$ ranges from 298--543~MeV/$c^2$ 
and the decay fractions range from 28--63\% and 
31--5\%, respectively.
The largest NR fraction obtained (31\%) 
remains substantially lower than that obtained without a  $\kappa$ 
resonance.

We have also studied the stability of our results 
with respect to the theoretical model.
For example, we modified the $\kappa$ Breit-Wigner to have a 
``running mass'' term as proposed by T\"ornqvist \cite{torn}.
We varied the momentum dependence of the 
$\kappa$ form factors.
We introduced Gaussian form factors for 
all other resonant states. We varied the shape of the NR term. 
In all cases we obtained similar results for the $\kappa$ mass 
and width within errors; however, the details of the parameterizations 
affect the relative $\kappa\pi^+$ and NR contributions by up to a factor of two. 

Finally, we have checked whether other models without a scalar $\kappa$ 
provide acceptable fits. 
We tried a toy model (T) by replacing the 
$\kappa$ complex Breit-Wigner by a Breit-Wigner amplitude with no phase 
variation. 
This model converged to a similar mass and width
($871\pm 10$ MeV/c$^2$ and $427\pm 23$ MeV/c$^2$, respectively) 
but with large decay fractions for this extra amplitude and for 
the NR amplitude, reflecting strong interference. 
The fast-MC gave $\langle \Delta w_{T,C}\rangle=60$ 
(rms of 16) for an ensemble generated according to Model~C, 
and $\langle \Delta w_{T,C}\rangle=-60$ for an ensemble
generated with toy model parameters. For the data, 
$\Delta w_{T,C}=45$; {\it i.e.}, the data prefers that the additional amplitude 
have a phase variation and not just a larger amplitude at low $ K \pi $ mass. 
We also replaced the scalar $\kappa$ resonance by vector and tensor 
resonances to test the angular distribution.  The vector resonance 
model (V) converged to mass and width values of $1103\pm 45$~MeV/$c^2$ 
and $350\pm 93$~MeV/$c^2$, respectively, with a decay fraction of only 1.8\% and 
a large NR fraction. The fast-MC gave $\langle \Delta w_{V,C}\rangle=140$ 
(rms of 23) for the ensemble generated according to Model~C, and 
$\langle \Delta w_{V,C}\rangle=-140$ 
for the ensemble generated with vector parameters. For the data, 
$\Delta w_{V,C}=116$; {\it i.e.}, the data prefers that the 
additional resonance be scalar rather than vector. We were 
not able to make the tensor model converge, the width being 
driven to large negative values. 
We also performed a variety of fits  to study 
the NR shape \cite{bediaga} in variants of Model A, {\em {\it i.e.}}, without 
an additional scalar amplitude. We fitted the NR amplitude 
to polynomials, and we also allowed for different interfering angular 
distributions, but none of these fits were as good as that of Model~C\@.

In summary, we have performed a Dalitz plot analysis of the 
decay $D^+\to K^-\pi^+\pi^+$. 
We compared models in which the signal 
amplitude $ {\cal A} $ is the coherent sum of a uniform NR
term and Breit-Wigner $ K \pi $ resonances. 
Our best fit is obtained when we include
an additional scalar resonance with a phase 
variation corresponding to that of a Breit-Wigner; this state
subsequently accounts for about half of the decay rate. 
The mass and width obtained are
$797\pm 19\pm 43$~MeV/c$^2$ and $410\pm 43\pm 87$~MeV/c$^2$,
respectively.
The fit mass and width of the $K^*_0(1430)$ depend on whether
this additional Breit-Wigner is included or not. When not included,
$m^{}_{K^*(1430)}=1416\pm 27$~MeV/c$^2$ and 
$\Gamma^{}_{K^*(1430)}=250\pm 21$~MeV/c$^2$
(statistical errors only), in agreement with PDG values \cite{pdg}. 
When included, 
$m^{}_{K^*(1430)}=1459\pm 7\pm 5$~MeV/c$^2$ and 
$\Gamma^{}_{K^*(1430)}=175\pm 12\pm 12$~MeV/c$^2$.
Overall we conclude that the scalar contribution to 
$ {\cal A} $ is not adequately described by the sum of a 
uniform non-resonant term and a $K^*_0(1430)$ term. 
Including an additional scalar 
resonance in ${\cal A}$ results in a good fit to the 
data while the mass and the width of the $K^*_0(1430)$
appear higher and narrower, respectively, than previous reported results.

We thank E.~van~Beveren and N.~T\"ornqvist for useful discussions. 
We gratefully acknowledge the assistance of the staffs of Fermilab and of all
the participating institutions.  This research was supported by CNPq (Brazil), 
CONACyT (Mexico), FAPEMIG (Brazil), the Israeli Academy of Sciences and Humanities, 
PEDECIBA (Uruguay), the U.S. Department of Energy, the U.S.-Israel Binational 
Science Foundation, and the U.S. National Science Foundation.

\bibliographystyle{unsrt}

\begin{thebibliography}{99.}

\bibitem{e691-kpipi} E691 Collaboration, J.C.~Anjos {\em et al.}, Phys. Rev.
D {\bf 48}, 56 (1993).
\bibitem{e687-kpipi} E687 Collaboration, P.L.~Frabetti {\em et al.}, 
Phys. Lett. B {\bf 331}, 217 (1994).
\bibitem{ref791} J.A.~Appel, Ann. Rev. Nucl. Part. Sci. {\bf 42}, 367 (1992);
D.~Summers {\em et al.}, hep-ex/0009015; S.\,Amato {\em et al.}, Nucl. Instrum. Methods 
A {\bf 324}, 535 (1993); E.M.~Aitala {\em et al.}, Eur. Phys. J. direct C {\bf 4}, 1 (1999).
\bibitem{cerenkov} D.~Bartlett {\em et al.}, Nucl. Instrum. Methods A {\bf 260}, 55 (1987).
\bibitem{pdg} D.E.~Groom {\em et al.}, Eur. Phys. Jour. C {\bf 15}, 1 (2000).
\bibitem{argus} ARGUS Collaboration, H. Albrecht  {\em{et al.}}, Phys. Lett. B
{\bf 308}, 435(1993); CLEO Collaboration, S.~Kopp {\em et al.}, Phys. Rev. D {\bf 63}, 
092001 (2001). 
\bibitem{blatt} J.M.~Blatt and V.F.~Weisskopf, Theoretical Nuclear Physics,
John Wiley \& Sons, New York, 1952.
\bibitem{d3pi} E791 Collaboration, E.M.~Aitala {\em et al.}, Phys. Rev. Lett. {\bf 86},
770 (2001).
\bibitem{torn} N.A.~T\"ornqvist, Z. Phys. C {\bf 68}, 647 (1995).
\bibitem{cleotau} CLEO Collaboration, D.M.~Asner {\em et al.}, Phys. Rev. D {\bf 61},
012002 (2000).
\bibitem{thresh} There remains the question about how best to characterize broad states 
near threshold. This subject is considered in Ref.~\cite{torn}, and also recently by 
E.~van~Beveren and G.~Rupp, Eur. Phys. J. C {\bf 22}, 493 (2001).
\bibitem{kapparef1} E.~van~Beveren {\em et al.}, Z. Phys. C {\bf 30}, 615 (1986);
S.~Ishida {\em et al.}, Prog. Theor. Phys. {\bf 98}, 621 (1997);
D.~Black {\em et al.}, Phys. Rev. D {\bf 58}, 054012 (1998);
J.A.~Oller {\em et al.}, Phys. Rev. D {\bf 59}
074001 (1999);  M.~Jamin {\em et al.}, Nucl. Phys. {\bf B587}, 331 (2000);
C.M.~Shakin and  H.~Wang, Phys. Rev. D {\bf 63}, 014019 (2001);
R.~Delbourgo and M.D.~Scadron, Int. J. Mod. Phys. A {\bf 13}, 657 (1998);
M.~Ishida, Prog. Theor. Phys. {\bf 101}, 661 (1999);
J.A.~Oller and E.~Oset, Phys. Rev. D {\bf 60}, 074023 (1999); 
F.E.~Close and
N.A.~T\"ornqvist, hep-ph/0204205 (2002).
\bibitem{nokapparef} A.V.~Anisovich and A.V.~Sarantsev, Phys. Lett. B {\bf 413}, 137 (1997);
S.N.~Cherry and M.R.~Pennington, Nucl. Phys. {\bf A688}, 823 (2001); N.A.~T\"ornqvist and A.D.~Polosa, in
{\em Heavy Quarks at Fixed Target}, edited by I.~Bediaga, J.~Miranda, and A.~Reis,  
Frascati Physics Series, Vol. XX (Laboratori Nazionali di Frascati, Roma, Italy, 2000), p. 385.
\bibitem{lass} LASS Collaboration, D.~Aston {\em et al.}, Nucl. Phys. {\bf B296}, 
493 (1988).
\bibitem{dunwoodie} A new fit to the LASS data for the $I=1/2$ S--wave 
$K\pi$ scattering amplitude in the elastic range (up to $K\eta'$ threshold) 
yields $m_{K^*_0(1430)}=1435\pm 5$ MeV/c$^2$ and 
$\Gamma_{K^*_0(1430)}=279\pm 6$ MeV/c$^2$. Private communication from W.M.~Dunwoodie for the
LASS Collaboration.
\bibitem{bediaga} I.~Bediaga, C.~G\"obel, and R.~M\'endez-Galain, Phys. Rev.
Lett. {\bf 78}, 22 (1997) and Phys. Rev. D {\bf 56}, 4268 (1997).
\end{thebibliography}

\end{document}